\documentclass[12pt]{article}

\usepackage{graphicx}
\usepackage{color}
\usepackage[colorlinks=true, urlcolor=blue, linkcolor=blue, 
citecolor=blue]{hyperref}
\usepackage{bm}
\usepackage{mathtext}
\usepackage{amssymb, amsmath}

\usepackage{epsfig}
\usepackage{bm}
\usepackage[usenames,dvipsnames]{xcolor}
\usepackage{color}

\begin{document}
\thispagestyle{empty}

\begin{center}

\textbf{\Large Star-forming galaxies as the origin of IceCube neutrinos: Reconciliation with Fermi-LAT gamma rays}

\vspace{50pt}
Sovan Chakraborty$^{a,b}$ and Ignacio Izaguirre$^{b}$
\vspace{16pt}



$^a$\textit{Indian Institute of Technology Guwahati,}\\
\textit{Guwahati - 781 039, Assam, India}\\
\vspace{16pt}

$^b$\textit{Max-Planck-Institut f\"ur Physik (Werner-Heisenberg-Institut),}\\
\textit{F\"ohringer Ring 6, 80805 M\"unchen, Germany}\\\vspace{3 mm}


\end{center}

\vspace{30pt}


\begin{abstract}

Cosmic ray accelerators like supernova and hypernova remnants in star forming galaxies are one of the most
plausible sources of the IceCube observed diffuse astrophysical neutrinos. The neutrino producing hadronic 
processes will also produce a diffuse gamma ray flux, constrained by the Fermi-LAT bounds. The fact that 
point sources like blazars also contribute to the diffuse gamma ray flux implies large gamma opacity 
of the neutrino sources. Indeed, for these high redshift star forming galaxies the gamma absorption during 
the intergalactic propagation can be significant. In addition, large gamma attenuation 
inside these extreme source galaxies can reduce the cascade component of the diffuse flux. 
Under the current astrophysical uncertainties affecting these absorptions processes,
we find the associated diffuse gamma ray flux can remain compatible with the current Fermi-LAT bounds.

\end{abstract}


\newpage
\section{Introduction}
\label{sec:intro}

The signature of high energy neutrinos in the IceCube (IC) experiment for the first time has opened the window 
observing neutrinos originating at the cosmic distances~\cite{Aartsen:2013bka,Aartsen:2013jdh,Aartsen:2014gkd,Aartsen:2014muf,Aartsen:2015knd,Aartsen:2015rwa}. The simple explanation of connecting these events to some high 
energy astrophysical point source remained elusive as the events are found to be compatible with an isotropic
distribution~\cite{Aartsen:2014gkd}. 
Although many astrophysical objects are not compatible with the observed IC flux, the catalog of 
proposed sources is nevertheless extensive. It ranges from long GRB remnants embedded in molecular clouds~\cite{Dado:2014mea} to 
active galactic nuclei (AGN)~\cite{Tamborra:2014xia} to intense star forming galaxies~\cite{He:2013cqa,Murase:2013rfa,Liu:2013wia,Chakraborty:2015sta,Senno:2015tra,Xiao:2016rvd}. 

These IC neutrino events observed at such high energies (TeV to PeV) have another important implication. 
One of the yet unanswered crucial questions of astroparticle physics is the nature of the interaction of the cosmic rays (CR).
The observation of these high energy neutrino events is indirectly pointing towards the hadronic interaction of the CR 
accelerators as neutrinos are expected to be produced through hadronic processes (via pions from pp or p$\gamma$ interactions).
This hadronic origin assumption of the neutrino flux will make the production of gamma rays from neutral pions unavoidable, 
creating a diffuse background of $\gamma$ ray flux. These multimessenger signatures are of great help to constrain the 
different astrophysical models compatible with IC high energy neutrino flux.
 
Among these possibilities there is one natural appealing candidate, consisting of stellar remnants (hypernova and supernovae 
remnants) inside galaxies with large star formation rate, known as star forming galaxies (SFGs). SFGs are distinguished between
normal star forming galaxies (NSFGs) and starburst galaxies (SBGs), i.e, galaxies with extraordinary star formation.
While hypernovae remnants (HNRs) are able to accelerate protons to hundreds of PeV energies thus generating PeV 
neutrinos, the more abundant supernovae remnants (SNRs) will dominate the flux in the hundred TeV range.
This scenario is not only able to explain the measured flux by IceCube, but also predicts a break on the neutrino spectrum around TeV 
energies~\cite{Chakraborty:2015sta}. Indeed, the large flux of neutrinos detected in the TeV energies with respect to the flux detected at 
PeV energies points to different sources of cosmic accelerators in the different energy ranges. 

The associated gamma ray flux with these sources will also lay in the GeV-TeV range. However, at this energy range, the 
intergalactic medium is highly opaque as the $\gamma$ rays interaction with the 
cosmic microwave background (CMB) and the extragalactic background light (EBL) remains significant. Of course, the gamma absorption 
depends on the distance traveled. SBGs, being distant objects, will result in a very efficient gamma rays absorption. Thus, the 
highest energy gamma rays cannot reach us from these far objects. However, this flux can reappear in the
lower energies. The initial $\gamma \gamma$ collisions in the intergalactic medium also produce a $e+$/$e-$ pairs, which will 
interact additionally to the EBL photons via the inverse Compton mechanism, resulting in a cascaded low energy $\gamma$ 
ray flux in the GeV energy range. Thus the total gamma ray flux in the GeV energies is one sensitive measure 
of all these different interaction processes. Measuring this diffuse $\gamma$ flux is one of the great challenges of the present day
physics experiments. One of the main goal of the Fermi-LAT collaboration experiment is to measure this isotropic diffuse gamma ray 
background (IGRB), in fact the measured extragalactic gamma ray background (EGB) includes the diffuse flux. The known blazar and other
$\gamma$ ray sources also contribute to the EGB. Any plausible hadronic model of the IC neutrinos should not overpopulate the bounds 
of the isotropic diffuse gamma ray background (IGRB)  measured by the Fermi-LAT collaboration~\cite{TheFermi-LAT:2015ykq}.
For a recent analysis of observed SFG-SBG point sources and their gamma ray-neutrino correlation, see~\cite{Moharana:2016mkl}.

The large number of neutrinos detected in the TeV energies have started putting pressure on this multimessenger 
picture as these larger fluxes demand a greater diffuse $\gamma$ background. Recently, the Fermi collaboration did a survey of the blazar sources in the high energy tail of the observed
Fermi EGB spectra~\cite{Ackermann:2014usa}. The stacking of all the blazar point sources gives the limit of diffuse gamma rays in GeV energies. 
In the energy interval of $50$ GeV to $1$ TeV this limit is found to be very close to the observed diffuse background, leaving no more than the 14\% percent of the observed fluxes 
to the hadronic channel gamma rays produced with neutrinos. This apparent tension in the picture is also pointing to a class of 
sources that are opaque in the gamma rays~\cite{Wang:2015mmh, Murase:2015xka} etc.
Recently, another study~\cite{Lisanti:2016jub} for these known gamma sources 
found that these limits can actually be more relaxed, i.e, 32 $\%$ non blazar gamma ray contribution. 
This is due to the fact that the percentage of the blazar component in EGB still has big uncertainties. 
However, there is no doubt that they own a significant fraction of the total EGB. Note that,
in the case of blazars, the leptonic models are found to be successful in explaining the source gamma spectra~\cite{LAT:2011aa}.
There is no co-production of neutrinos in these models. Of course, invoking a hadronic model with subsequent neutrino production for the blazar gammas could relieve this tension.

In the following, we focus on the SBGs and calculate the limits 
coming from the different components of the diffuse gamma and neutrino spectra. The large dimensions of the early galaxies 
with large number of background photons make sure that the gamma rays with higher energies (1-10 TeV) cannot escape. 
Thus the cascade component of the gamma rays coming from the high energy part can get greatly attenuated. The lower energy 
component with ordinary intergalactic absorption can still reach the detectors.
However, the flux depends on several free parameters and the estimation of the EBL. In this work, we try to understand the 
broad picture, independent of the parameter uncertainties. We find that the neutrino flux from this model explaining the IC 
events can still be accommodated in view of the recent Fermi-LAT non-blazar EGB estimate. Under the present uncertainties of the 
EBL and considering the intragalactic absorption, the hadronic gammas can remain under the observed 
non-blazar EGB gamma limit. Thus not only the parameter space describing this model can get narrowed down but also the absorption 
models of gamma rays can get constrained. \cite{TheFermi-LAT:2015ykq,Bechtol:2015uqb}. 

The outline of this paper is the following. In section~\ref{sec:flux} we briefly describe the $\nu$ and the $\gamma$ flux production 
by the stellar remnants (SRs) embedded in the star forming galaxies. In section~\ref{sec:diffga} we describe the details of gamma 
ray spectrum of SFGs, both the cascading and non cascading component. Then, section~\ref{sec:intra} , describes the absorption of the $\gamma$ rays 
inside the source galaxies and their effect on the total diffuse gamma ray flux. 
In section~\ref{sec:gamma} we discuss the different uncertainties in the estimation of the gamma spectra.
Finally, we comment and conclude in section ~\ref{sec:conclus}.

\section{Diffuse $\nu$ and $\gamma$ flux from stellar remnants}
\label{sec:flux}

The observed diffuse flux ($\frac{dN(E_{\alpha}^{ob})}{dE_{\alpha}^{ob}}, \alpha = \nu, \gamma$) from each kind of stellar remnants
embedded in the star forming galaxy is ,
\begin{equation}
\frac{dN(E_{\alpha}^{ob})}{dE_{\alpha}^{ob}} = \frac{c}{4 \pi H_0} \int_{0}^{z_{max}} \frac{dN (E_{\alpha})}{dE_{\alpha}}
\frac{R_{SR}(z)~dz}{\sqrt{\Omega_M(1+z)^3+\Omega_\lambda}}\,,
\label{eq:DSRalpha}
\end{equation}

where the $R_{\text{SR}}(z)$ is the stellar remnant (SNR or HNR) rate with $dN (E_{\alpha})/dE_{\alpha}$ being the flux at the 
source and ($E_{\alpha}^{\text{ob}} = E_{\alpha}/(1+z)$) is the observed energy for source energy $E_{\alpha}$. For the Hubble 
parameter ($H_{0}$), the matter energy density ($\Omega_{\text{M}}$) and the dark energy density 
($\Omega_{\lambda}$)  we use 0.69 $\text{km}~\text{s}^{-1}\text{Mpc}^{-1}$, 0.27 and 0.73, respectively. 

Depending on the nature of the host galaxy, $\textit{i.e.}$ NSFG or SBG, the source flux $dN (E_{\alpha})/dE_{\alpha}$ will 
be different. The total flux is a weighted sum of the contributions from SBGs and NSFGs
and also of the different kinds of remnant population, HNR and SNR. The relative population of the SBGs with respect to NSFGs  ($f_{\text{SBG}}$) is estimated to be 
 $10$--$20$ percent \cite{Rodighiero:2011px,Lamastra:2013lfp,Gruppioni:2013jna}. 
The source flux from a particular type of host galaxy is given by ~\cite{Kelner:2006tc},
\begin{equation}
\frac{dN(E_{\alpha})}{dE_{\alpha}} = \int_{E_{\alpha}}^{\infty} \frac{\eta_{\pi}(E_p)}{\kappa} J_{p}(E_p) 
F_{\alpha}(\frac{E_{\alpha}}{E_p},E_p) 
\frac{dE_p}{E_p}\,,
\label{eq:SRNalpha}
\end{equation} 
with $\kappa$ (0.2) being the elasticity and $\eta_{\pi}$ the efficiency of the pion production. The primary 
proton spectrum, $J_{p}$ $\sim E_p^{2} \exp (-E_p/E_p^{max})$,  depends on the maximum proton energy $E_p^{max}$. The 
normalization of $J_{p}$ comes from the total proton energy ($E_p^{\text{T}}$) which is a  fraction of the total ejected energy of 
the stellar remnant. Both the $E_p^{max}$ and the total proton energy estimates the energetics of the stellar remnant, for the 
HNRs $E_p^{\text{T}}$ is in the range $5\times 10^{51}$--$10^{52}$ erg~\cite{Kulkarni:1998qk}, whereas for SNRs the $E_p^{\text{T}}$ 
is expected to be at least one order lower. The maximum energy ($\text{E}_\text{p}^{\text{max}}$) for the SNRs and HNRs populations 
are  in the range 1--10 PeV and $10^2$--$10^3$ PeV, respectively~\cite{Gaisser,Wang:2007ya}.

The pion production efficiency ($\eta_{\pi}$) gives a measure of the escape time of cosmic ray protons in the host galaxy, 
i.e. if the proton confinement time is long enough to collide with the interstellar medium (ISM) gas and loose energy via 
pion and other subsequent secondary particle production. The SBGs with their larger proton density~\cite{Tacconi:2005nx} are 
expected to have more efficient energy loss compare to the NSFGs producing larger secondary neutrino and gamma ray fluxes. 

Finally, the total secondary background also depends on the relative population of the different stellar remnants and 
their redshift dependence . The $R_{\text{SR}}(z)$ is considered to follow the star formation
rate $R_{\text{SFR}}(z)$~\cite{Hopkins:2006bw,Yuksel:2006qb}. In particular, the SNR rate is $R_{\text{SNR}}(z)= 1.22 \times 10^{-2} R_{\text{SFR}}(z)
\text{M}_{\odot}^{-1}$~\cite{Baldry:2003xi} and the HNR rate is
$R_{\text{HNR}}(z) \leq 10^{-4} R_{\text{SFR}}(z) \text{M}_{\odot}^{-1}$~\cite{Guetta:2006gq,Wanderman10,Bhattacharjee:2007gc} . 

Thus both the secondary neutrino and gamma ray characteristics are dependent on several source factors; the energetics, mass 
and explosion mechanism of the stellar remnants and the host galaxies molecular densities,
magnetic field and dimensions. The neutrinos once emitted from the source galaxies travel freely but the gamma rays
gets absorbed both inside the galaxy and during intergalactic path. In the following, we explain this gamma ray absorption in more detail. 

\section{Gamma ray absorption in the intergalactic medium}
\label{sec:diffga}
The emitted gamma rays, while traveling through intergalactic medium, interact with the photons of EBL and CMB ( depending on their energy) and 
get absorbed through $e+$/$e-$ pair production. The higher energy photons are more susceptible to the interaction as the absorption probability increases 
with energy and distance of the gamma ray source.
Therefore the diffuse gamma ray background mostly gets populated by the low energy, unabsorbed component.\
In particular, the optical depth $\tau_{\gamma \gamma} (E_{\gamma}, z)$ of the photons depend 
on both the photon energy and the source redshift. For example, only nearby ($z < 1$) emissions will allow photons 
above 1 TeV to propagate. In fact, even from nearby distances ($z < 0.1$) only photons up to tens of TeV can travel without being attenuated. 

The efficiency of this absorption strongly depend on the distribution of the EBL in the intergalactic media. 
There are large uncertainties regarding the EBL estimates and thus also on the universe's opacity. 
These similarly imply uncertainties in the diffuse gamma spectra. 
In this section, we will focus only on one such EBL estimate~\cite{Stecker:2016fsg} and demonstrate the order of the resultant flux.
However,  to generate a broader understanding of the topic, in section~\ref{sec:gamma},
we will look into different estimations of the optical depth~\cite{Stecker:2016fsg, Finke:2009xi}.

The diffuse gamma flux have another low energy contribution coming from the cascading of the 
fluxes with energies higher than 10 TeV. The $e+$/$e-$ pair scatter with the EBL photons via inverse
Compton mechanism producing the electromagnetic cascade. This process runs till the cascaded photons are 
above the $e+$/$e-$ pair production threshold. 
The exact details of these processes are complex and beyond the scope of this work~\cite{Berezinsky:1975zz}. However, when this cascading process 
is sufficiently
developed, if adopts a near-universal form, namely~\cite{Berezinsky:1975zz,Coppi:1996ze}
\begin{equation}
G_{E_\gamma} \propto 
 \begin{cases}
    (E_\gamma/E^{br}_\gamma)^{-1/2},&  (E_\gamma\leq E^{br}_\gamma)\\
    (E_\gamma/E^{br}_\gamma)^{1-\beta},              & (E^{br}_\gamma<E_\gamma\leq E^{cut}_\gamma),
\end{cases}
\end{equation}

where $E_{\gamma}G_{E_\gamma}$ is the shape of the gamma ray spectrum produced at the source.
$E^{cut}_{\gamma}$ is given by the intergalactic suppression due to pair creation, which depends on the distance $z$ of the source, the energy 
$E_\gamma$ and on the specific cosmic opacity estimate chosen. The parameter $E^{br}_\gamma$ is given by 
$E^{br}_\gamma\simeq0.034(\text{GeV})(E^{cut}_{\gamma}/0.1\text{TeV})^2((1+z)/2)^2$, and $\beta$ will be taken to be 2~\cite{Berezinsky:1975zz,Coppi:1996ze}.

\begin{figure}[!t]
\begin{center}
 \includegraphics[angle=0,width=0.8\textwidth]{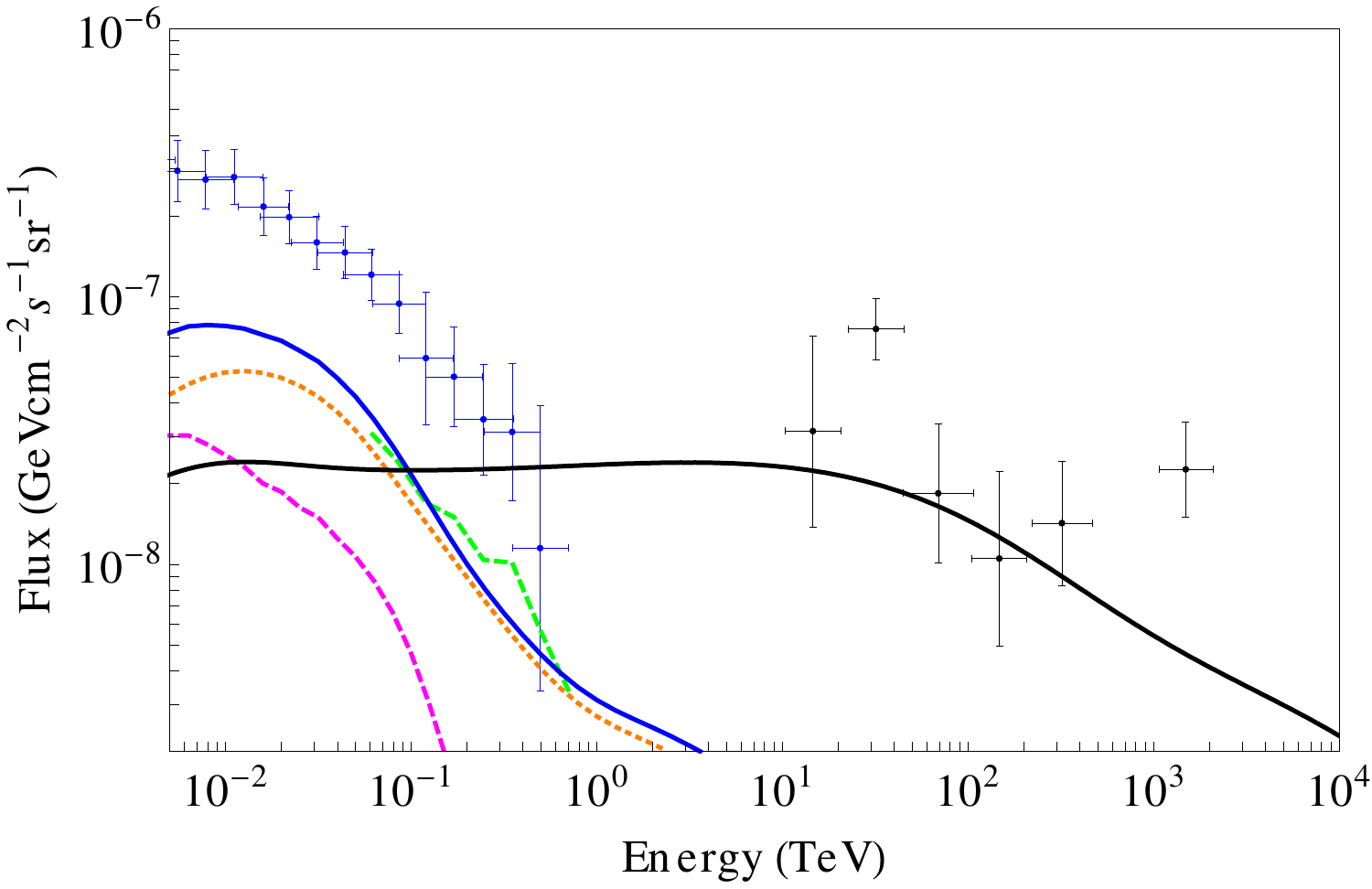}
\end{center}
\caption{ 
The single flavor diffuse neutrino spectra and the IC data~\cite{Aartsen:2015knd} are given by 
the black continuous curve and points, respectively. The blue continuous line is the total gamma flux when using the ~\cite{Stecker:2016fsg} estimation for EBL absorption. The pink, dashed curve is the cascading counterpart and the dotted orange line is the usual unabsorbed counterpart.
The blue points are the IGRB measured by Fermi-LAT, and the green, dashed green line is the 14 $\%$ non-blazar limit for the EGB~\cite{Ackermann:2014usa}.}
\label{fig-intergalacticAbs}
\end{figure}

In figure~\ref{fig-intergalacticAbs}, we show the `multimessenger spectra' produced by this scenario.
The black data points describe the IC observed flux~\cite{Aartsen:2015knd}, and the solid, black line is the neutrino flux obtained with the above discussed model. 
For this particular curve, we have assumed 20 $\%$ of SBGs, and a SNR and HNR remnant luminosity of $5\times10^{51}$~erg/s and $10^{52}$erg/s, 
respectively.
This neutrino diffuse flux is compatible with the experimental points. In fact, we normalize our neutrino flux with the IC best fit at 100 TeV~\cite{Aartsen:2015knd}.

The blue data points on the figure are the Fermi-LAT data points for the IGRB~\cite{TheFermi-LAT:2015ykq}. 
On the other hand, according to \cite{Ackermann:2014usa} at least 86 $\%$ of the extra-galactic diffuse gamma-ray background (EGB) are due to known point 
sources like blazar, GRB etc. 
The rest 14 $\%$ is the upper limit of the gamma flux correlated to the neutrino spectra.
The green, dashed line is the lower limit of the best-fit 14 $\%$ non-blazar emission in the 0.05-1 TeV EGB~\cite{Ackermann:2014usa}.
The dotted orange curve represents the unabsorbed diffuse gamma ray flux associated to the neutrino flux, using the EBL estimates
provided by~\cite{Stecker:2016fsg}. The dashed pink line is the diffuse flux 
from the cascading component of the original gamma flux. The total gamma flux is given by the blue continuous curve. 
The flux above the hundred GeV range is primarily determined by the intergalactic absorption,
as we go to lower energies (GeV), the cascading part will also have a significant contribution.

It is evident that the diffuse gamma ray flux is overpopulating the 14 $\%$ limit and can be thought of 
as an evidence against SFG origin of the IceCube neutrinos~\cite{Bechtol:2015uqb}. Off course, as we have already pointed out that these results 
depend on the intergalactic gamma ray opacity model, 
which will be discussed in section~\ref{sec:gamma}. However, there is another crucial factor missing in these estimations. 
In between the two components of the SFGs, i.e, NSFGs and SBGs, the dominant contribution of both the $\nu$ and $\gamma$ fluxes are coming from the SBGs.
The SBGs with their larger photon density, length scale and magnetic field strength have significantly larger gamma ray opacity compared to the NSFGs. 
This gamma absorption in these SBGs can give rise to interesting scenarios and in the next section
we discuss this feature in detail.

\section{Gamma ray absorption inside the SBGs}
\label{sec:intra}

The large photon number density present in the interior of SBGs makes this environment opaque to TeV-PeV gamma rays~\cite{Inoue :2010,Lacki :2013}. This will imply that highly energetic gamma rays will interact inside the galaxy
despite of its small scale when compared to intergalactic distances and will create  $e$+/$e-$ pairs. However, unlike the case of the intergalactic medium, the larger 
magnetic fields will change the fate
of the low energy cascade component in SBGs. The $e$+/$e-$ pairs produced in these collisions will loose their energy through synchrotron radiation, resulting in emission of photons with energy
below GeV and therefore inhibiting the cascading~\cite{Chang:2014hua}. These models are based on the observation of the nearby ($z\simeq10^{-3} $) SBGs~\cite{Inoue :2010,Torres:2004ui,DomingoSantamaria:2005qk}. 
These phenomenological models of SBGs have been extrapolated to the average properties of the far away starburst galaxies with radius of the order of 1 kpc and 
intense magnetic fields (B~$\sim$~mG)~\cite{Thompson:2006is,Lacki:2009mj}. Of course, a part of the gamma rays will escape the galaxy without interacting and contribute 
to the cascading. However, this fraction is found to be negligible.  
Based on reference~\cite{Chang:2014hua}, for a SBG of radius 1 kpc more than 90 $\%$ of the produced gamma rays convert into synchrotron radiation. Indeed, this significantly reduce the cascading counterpart of the diffuse gamma ray flux.

\begin{figure}[!t]
\begin{center}
 \includegraphics[angle=0,width=0.8\textwidth]{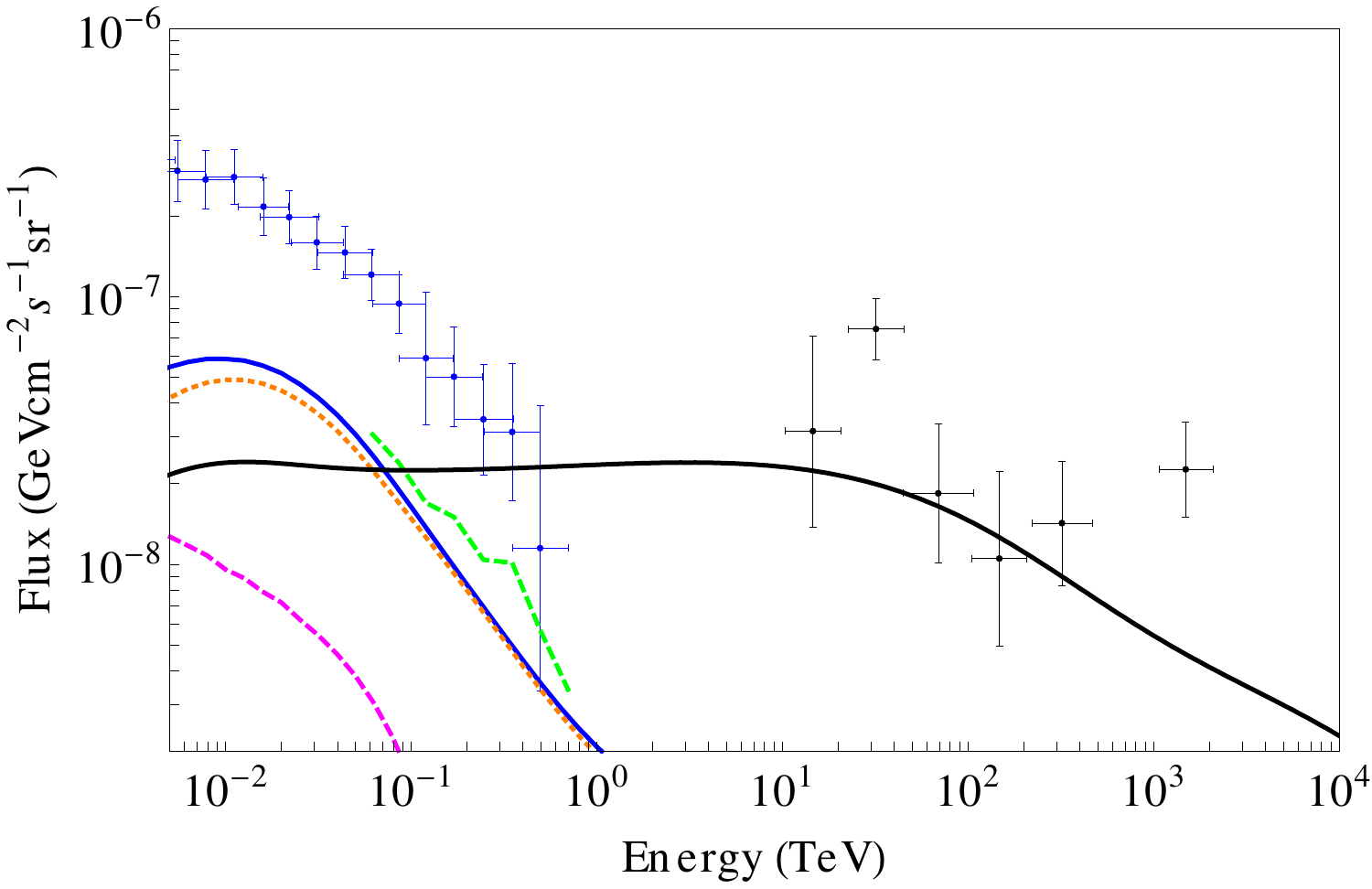}
\end{center}
\caption{ Same curve legends as in the figure~\ref{fig-intergalacticAbs}. However, the gamma fluxes are now estimated with the intragalactic absorption in the SBGs~\cite{Torres:2004ui}.}
\label{fig-internalAbs}
\end{figure}

The figure~\ref{fig-internalAbs} describes this effect of the intragalactic gamma ray absorption. The neutrino 
diffuse flux is the same as in figure~\ref{fig-intergalacticAbs} and the same estimation~\cite{Stecker:2016fsg} of the  
intergalactic absorption has been adopted. The blue data points are again the IGRB flux measured by Fermi-LAT.
The orange, dotted curve represents the flux which has not been absorbed in the intergalactic medium and the new cascade flux is shown by the orange dashed line. 
The total diffuse gamma ray flux is described by the blue, continuous curve. From the figure, one can appreciate the effect on the cascade flux and therefore on the total gamma ray flux. 
More precisely, comparing the cascade component with that of the figure~\ref{fig-intergalacticAbs} shows a significant reduction in the flux due to the absorption inside the SBGs. 
This crucial feature, which has been often overlooked, is indeed the main reason why the gamma ray flux can avoid overpopulating the non-blazar EGB bound~\cite{Ackermann:2014usa}.

\section{Sources of uncertainties}
\label{sec:gamma}
The spectra of both the neutrino and gamma ray depend on several model parameters with large uncertainties. The flux estimations with the present knowledge of the astrophysical parameters give 
rise to a band rather than a single line~\cite{Chakraborty:2015sta}. In section.~\ref{sec:flux} the neutrino flux uncertainties from the astrophysical sources has been discussed in detail. 
The gamma ray counterpart not only get influenced by these source uncertainties but also 
from the model dependence of their propagation inside the galaxy and intergalactic medium.

The gamma absorption in the intergalactic medium can get significant influence from the deviations in the EBL estimations which in turn depend on the uncertainties of the observational data from deep galaxy surveys.
The curves shown in the figure~\ref{fig-intergalacticAbs},~\ref{fig-internalAbs} are done with the empirical determination of the intergalactic background light of reference~\cite{Stecker:2016fsg}. 
However, there are several other estimations of the opacity or optical depths ~\cite{Kneiske:2003tx,Gilmore:2011ks,Franceschini:2008tp} of the intergalactic medium.  These differences actually generate from systematic and statistical 
uncertainties of different experiments and astrophysical sources. 

\begin{figure}[!t]
\begin{center}
 \includegraphics[angle=0,width=0.8\textwidth]{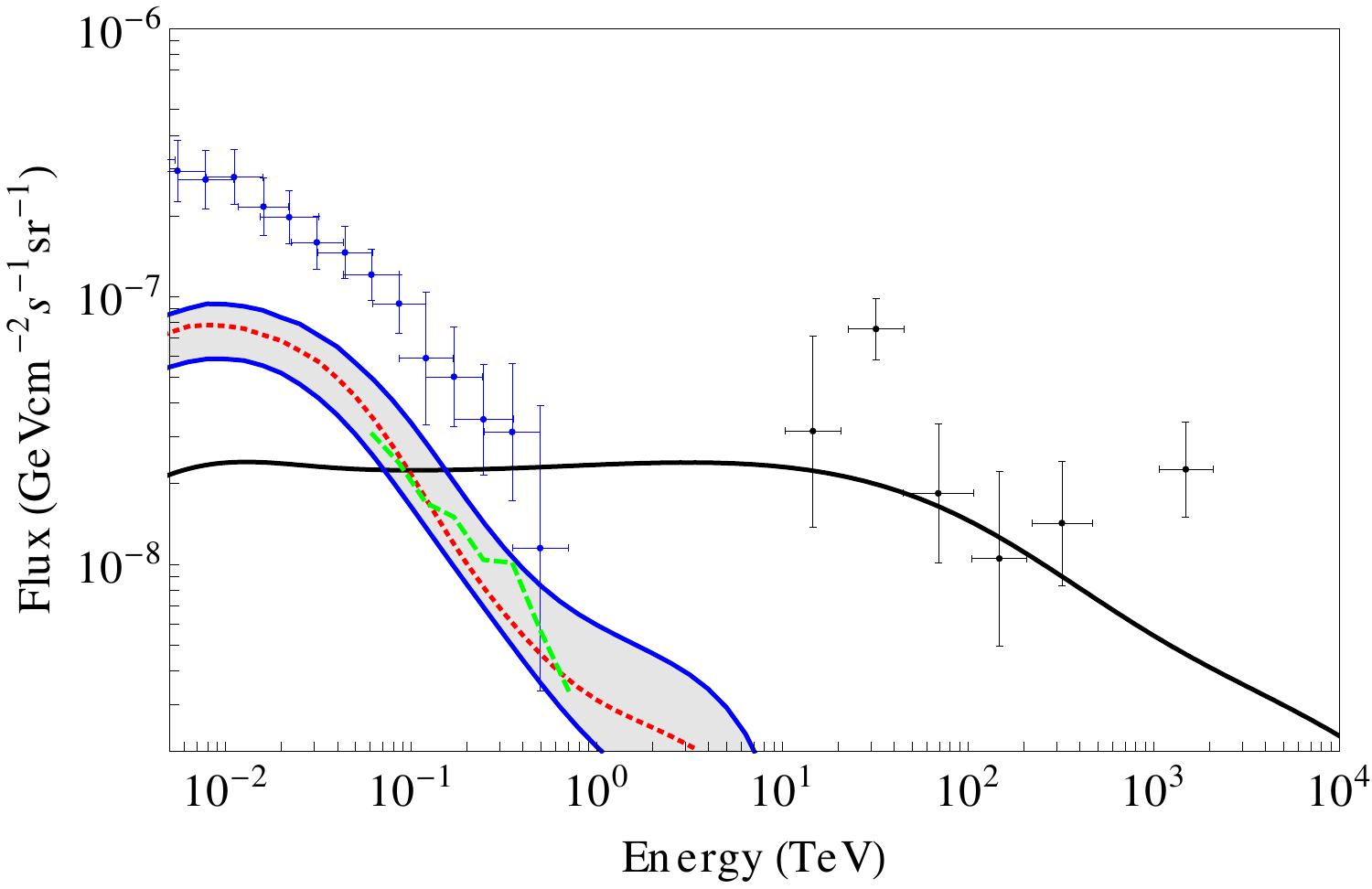}
\end{center}
\caption{ The IC and Fermi-LAT data points are the same as the figure~\ref{fig-intergalacticAbs} and the green dashed line is the non-blazar EGB limit. 
The black continuous line is the neutrino flux line. The corresponding gamma rays for the~\cite{Finke:2009xi} is the red dotted curve, the Grey band between 
the blue lines are from the~\cite{Stecker:2016fsg} models. The gamma fluxes are with both intergalactic and intragalactic absorption. }
\label{fig-gammaband}
\end{figure}

The main motivation of this discussion is to gain a general understanding of the problem rather than fixing details of the models. 
Therefore, we use two extreme estimations for the intergalactic absorption from~\cite{Stecker:2016fsg}, one with the largest opacity and another one with the minimal opacity~\cite{Stecker:2016fsg}.
These two curves give us the upper and lower limit of the band of the uncertainties coming from opacities. We also estimate the flux with another fit of the optical depth~\cite{Finke:2009xi}.
In figure~\ref{fig-gammaband} the green dashed line is the limit from the 14 $\%$ of the EGB, allowed for non-blazar sources. The IC and Fermi-LAT data points are the same as in the figure~\ref{fig-intergalacticAbs}.
The black continuous line is the neutrino flux line.
The corresponding gamma rays for the~\cite{Finke:2009xi} is the red dotted curve, the gray band between the blue lines are from the~\cite{Stecker:2016fsg} extreme cases.
The gamma fluxes are with both intergalactic, intragalactic absorption and also with the cascade component.

The gamma band shows the uncertainties coming from the intergalactic gamma opacity can be significant, however within these model
uncertainties the tension~\cite{Ackermann:2014usa} in the gamma and neutrino correlation does not seem to be very significant. 
The SBG models can still successfully explain the IC flux and do not overpopulate the gamma bounds. Off course, the uncertainties in the neutrino spectra can give rise to more possibilities. 

Another important source of uncertainty arises from the intragalactic absorption. The optical depths in primitive SBGs are difficult 
to measure. The size and shape of these high redshift SBGs, i.e, the absorption path length of the gamma rays are another source of the uncertainty. 
Moreover, the main reason for the suppression of cascade gamma component in SBGs is their large magnetic fields which increases the synchrotron radiation 
from the $e$+/$e-$ pairs instead of populating diffuse gamma rays above GeV energy.
However, the observation of the nearby SBGs~\cite{Inoue :2010, Torres:2004ui,DomingoSantamaria:2005qk} are the only direct source of information about their properties. 
The high redshift primitive galaxies are expected to be more extreme in size, density and magnetic field strength than the observed local ones.
In this sense, our estimation of the intragalactic absorption is conservative. The diffuse gamma flux component from the SBGs might be smaller compare to our estimations. 
Also note that this absorption process is not significant for the NSFGs, so an increase in the NSFG fraction can still increase the cascade component.

There are proposals that the apparent tension between the diffuse neutrino flux in IC and the IGRB in Fermi-LAT might be pointing towards a different redshift evolution of the SBGs. 
This might be a plausible scenario however that might not point towards a sudden cut off in the redshift evolution of the SBGs~\cite{Chang:2016ljk}. 
Such a solution is more like a fine tuned scenario, rather a detailed look at the connection of different SBG models and their overall redshift evolution would 
be interesting~\cite{Murase:2016gly}. Anyway, our estimation shows that with the present astrophysical knowledge and uncertainties the 
tension between IC neutrinos and IGRB is not significant.

\section{Conclusions}
\label{sec:conclus}

Stellar remnants in normal star forming galaxies and star burst galaxies have been proven to provide a diffuse neutrino flux compatible with the results obtained in the IceCube experiment. 
The existence of two different kind of accelerators 
SNRs and HNRs in those galaxies also give rise to the characteristic break in the neutrino spectrum. Nevertheless, because of the hadronic origin of the neutrino flux associated gamma ray production would be unavoidable. 
In light of the recent studies of the diffuse gamma spectra of Fermi-LAT and the blazar contribution to this diffuse spectra a tension between the diffuse gamma and neutrino spectra has been suggested. 
We have studied how the different models for the intergalactic absorption affect the diffuse gamma ray spectrum. Moreover, very importantly we have added the internal absorption of gamma rays present in SBGs 
to show how this modify the spectrum in the GeV-TeV energies. In particular, the cascading part of the diffuse gamma spectra gets a major reduction from this galactic absorption. Combining these two effects, 
we have shown that under reasonable assumption of the astrophysical uncertainties this diffuse $\gamma$ ray flux can be compatible 
with the current experimental limits.
The main conclusion from this study is therefore that the diffuse gamma ray flux associated to this SBG models cannot be used to rule out this scenario that stellar remnants embedded in SBGs are the main 
source of the diffuse IC neutrino spectra. This work is a reassurance of this previously presented models. 

Therefore, the neutrino diffuse flux spectrum footprint predicting a spectra break around 100 TeV is a better tool to test these models describing the origin of the cosmic neutrinos. 
This is based on the fact that the weakly interacting neutrinos do not suffer from the above mentioned absorptions and one can avoid the uncertainties introduced by the EBL density and the SBGs internal absorption. 
Indeed, it would be interesting to investigate the IceCube data accumulation time frame and the statistical significance necessary to confirm or to rule out this spectral break scenario. 
Nevertheless, as shown in this work, it is always essential to provide a consistent multimessenger picture.

\section*{Acknowledgments}
II thanks Marcel Strzys and David Paneque for fruitful discussions and pointing out interesting ideas. We thank Edoardo Vitagliano, Georg Raffelt and Irene Tamborra for useful discussions.
S.C.\ acknowledges support from the European Union through a Marie Curie Fellowship, Grant No.\ PIIF-GA-2011-299861 and through the ITN “Invisibles”, 
Grant No.\ PITN-GA-2011-289442 in the initial stage of this work and the Max-Planck-Institut f\"ur Physik, Munich for hospitality during the final stage of this work.


\begin{thebibliography}{99} 

\bibitem{Aartsen:2013bka}
  M.~G.~Aartsen {\it et al.} [IceCube Collaboration],
  ``First observation of PeV-energy neutrinos with IceCube,''
  Phys.\ Rev.\ Lett.\  {\bf 111} (2013) 021103
  [arXiv:1304.5356 [astro-ph.HE]].


\bibitem{Aartsen:2013jdh}
  M.~G.~Aartsen {\it et al.}  [IceCube Collaboration],
  ``Evidence for High-Energy Extraterrestrial Neutrinos at the IceCube Detector,''
  Science {\bf 342} (2013) 1242856
  [arXiv:1311.5238 [astro-ph.HE]].

\bibitem{Aartsen:2014gkd}
  M.~G.~Aartsen {\it et al.}  [IceCube Collaboration],
  ``Observation of High-Energy Astrophysical Neutrinos in Three Years of IceCube Data,''
  Phys.\ Rev.\ Lett.\  {\bf 113} (2014) 101101
  [arXiv:1405.5303 [astro-ph.HE]].
  
  
\bibitem{Aartsen:2014muf}
  M.~G.~Aartsen {\it et al.}  [IceCube Collaboration],
  ``Atmospheric and astrophysical neutrinos above 1~TeV interacting in IceCube,''
   Phys.\ Rev.\ D {\bf 91}  (2015) 022001
    [arXiv:1410.1749 [astro-ph.HE]].
    
    
\bibitem{Aartsen:2015knd}
  M.~G.~Aartsen {\it et al.} [IceCube Collaboration],
  ``A combined maximum-likelihood analysis of the high-energy astrophysical neutrino flux measured with IceCube,''
  Astrophys.\ J.\  {\bf 809} (2015) no.1,  98
  [arXiv:1507.03991 [astro-ph.HE]].
    
\bibitem{Aartsen:2015rwa}
  M.~G.~Aartsen {\it et al.} [IceCube Collaboration],
  ``Evidence for Astrophysical Muon Neutrinos from the Northern Sky with IceCube,''
  Phys.\ Rev.\ Lett.\  {\bf 115} (2015) no.8,  081102
  [arXiv:1507.04005 [astro-ph.HE]].
    

\bibitem{Dado:2014mea}
  S.~Dado and A.~Dar,
  ``Origin of the High Energy Cosmic Neutrino Background,''
  Phys.\ Rev.\ Lett.\  {\bf 113} (2014) 19,  191102
  [arXiv:1405.5487 [astro-ph.HE]].
 
\bibitem{Tamborra:2014xia}
  I.~Tamborra, S.~Ando and K.~Murase,
  ``Star-forming galaxies as the origin of diffuse high-energy backgrounds: Gamma-ray and neutrino connections, and implications for starburst history,''
  JCAP {\bf 1409} (2014) 09,  043
  [arXiv:1404.1189 [astro-ph.HE]].

\bibitem{He:2013cqa}
  H.~N.~He, T.~Wang, Y.~Z.~Fan, S.~M.~Liu and D.~M.~Wei,
  ``Diffuse PeV neutrino emission from ultraluminous infrared galaxies,''
  Phys.\ Rev.\ D {\bf 87} (2013) 6,  063011
  [arXiv:1303.1253 [astro-ph.HE]].
  
\bibitem{Murase:2013rfa}
  K.~Murase, M.~Ahlers and B.~C.~Lacki,
  ``Testing the Hadronuclear Origin of PeV Neutrinos Observed with IceCube,''
  Phys.\ Rev.\ D {\bf 88} (2013) 12,  121301
  [arXiv:1306.3417 [astro-ph.HE]].

\bibitem{Liu:2013wia}
  R.~Y.~Liu, X.~Y.~Wang, S.~Inoue, R.~Crocker and F.~Aharonian,
  ``Diffuse PeV neutrinos from EeV cosmic ray sources: semi-relativistic hypernova remnants in star-forming galaxies,''
  Phys.\ Rev.\ D {\bf 89} (2014) 083004
  [arXiv:1310.1263 [astro-ph.HE]].
  
  
\bibitem{Chakraborty:2015sta}
  S.~Chakraborty and I.~Izaguirre,
  ``Diffuse neutrinos from extragalactic supernova remnants: Dominating the 100 TeV IceCube flux,''
  Phys.\ Lett.\ B {\bf 745} (2015) 35
  [arXiv:1501.02615 [hep-ph]].

\bibitem{Senno:2015tra}
  N.~Senno, P.~Mészáros, K.~Murase, P.~Baerwald and M.~J.~Rees,
  ``Extragalactic star-forming galaxies with hypernovae and supernovae as high-energy neutrino and gamma-ray sources: the case of the 10 TeV neutrino data,''
  arXiv:1501.04934 [astro-ph.HE].  

\bibitem{Xiao:2016rvd}
  D.~Xiao, P.~Mészáros, K.~Murase and Z.~g.~Dai,
  ``Revisiting the Contributions of Supernova and Hypernova Remnants to the Diffuse High-Energy Backgrounds: Constraints on Very-High-Redshift Injections,''
  arXiv:1604.08131 [astro-ph.HE].
    
    
\bibitem{TheFermi-LAT:2015ykq}
  M.~Ackermann {\it et al.} [Fermi-LAT Collaboration],
  ``Resolving the Extragalactic $\gamma$-ray Background above 50 GeV with Fermi-LAT,''
  arXiv:1511.00693 [astro-ph.CO].
 
  
\bibitem{Ackermann:2014usa}
  M.~Ackermann {\it et al.} [Fermi-LAT Collaboration],
  ``The spectrum of isotropic diffuse gamma-ray emission between 100 MeV and 820 GeV,''
  Astrophys.\ J.\  {\bf 799} (2015) 86
  [arXiv:1410.3696 [astro-ph.HE]].
   
\bibitem{Moharana:2016mkl}
  R.~Moharana and S.~Razzaque,
  arXiv:1606.04420 [astro-ph.HE].

\bibitem{Wang:2015mmh}
  X.~Y.~Wang and R.~Y.~Liu,
  ``Tidal disruption jets of supermassive black holes as hidden sources of cosmic rays: explaining the IceCube TeV-PeV neutrinos,''
  Phys.\ Rev.\ D {\bf 93} (2016) no.8,  083005
  [arXiv:1512.08596 [astro-ph.HE]].

\bibitem{Murase:2015xka}
  K.~Murase, D.~Guetta and M.~Ahlers,
  Phys.\ Rev.\ Lett.\  {\bf 116} (2016) no.7,  071101
  doi:10.1103/PhysRevLett.116.071101
  [arXiv:1509.00805 [astro-ph.HE]].

\bibitem{Lisanti:2016jub}
  M.~Lisanti, S.~Mishra-Sharma, L.~Necib and B.~R.~Safdi,
  arXiv:1606.04101 [astro-ph.HE].

\bibitem{LAT:2011aa}
  A.~A.~Abdo {\it et al.} [LAT and MAGIC Collaborations],
  ``Fermi large area telescope observations of Markarian 421: The missing piece of its spectral energy distribution,''
  Astrophys.\ J.\  {\bf 736} (2011) 131
  [arXiv:1106.1348 [astro-ph.HE]].
  
  
\bibitem{Bechtol:2015uqb}
  K.~Bechtol, M.~Ahlers, M.~Di Mauro, M.~Ajello and J.~Vandenbroucke,
  ``Evidence against star-forming galaxies as the dominant source of IceCube neutrinos,''
  arXiv:1511.00688 [astro-ph.HE].

\bibitem{Rodighiero:2011px}
  G.~Rodighiero, E.~Daddi, I.~Baronchelli, A.~Cimatti, A.~Renzini, H.~Aussel, P.~Popesso and D.~Lutz {\it et al.},
  ``The lesser role of starbursts for star formation at z=2,''
  Astrophys.\ J.\  {\bf 739} (2011) L40
  [arXiv:1108.0933 [astro-ph.CO]].

\bibitem{Lamastra:2013lfp}
  A.~Lamastra, N.~Menci, F.~Fiore and P.~Santini,
  ``The interaction-driven starburst contribution to the cosmic star formation rate density,''
  Astron.\ Astrophys.\  {\bf 552} (2013) A44
  [arXiv:1302.1363 [astro-ph.CO]].
  
\bibitem{Gruppioni:2013jna}
  C.~Gruppioni, F.~Pozzi, G.~Rodighiero, I.~Delvecchio, S.~Berta, L.~Pozzetti, G.~Zamorani and P.~Andreani {\it et al.},
  ``The Herschel PEP/HerMES Luminosity Function. I: Probing the Evolution of PACS selected Galaxies to z~4,''
  Mon.\ Not.\ Roy.\ Astron.\ Soc.\  {\bf 432} (2013) 23
  [arXiv:1302.5209 [astro-ph.CO]].

\bibitem{Kelner:2006tc}
  S.~R.~Kelner, F.~A.~Aharonian and V.~V.~Bugayov,
  ``Energy spectra of gamma-rays, electrons and neutrinos produced at proton-proton interactions in the very high energy regime,''
  Phys.\ Rev.\ D {\bf 74} (2006) 034018
   [Erratum-ibid.\ D {\bf 79} (2009) 039901]
  [astro-ph/0606058].
 
  
\bibitem{Kulkarni:1998qk}
  S.~R.~Kulkarni, D.~A.~Frail, M.~H.~Wieringa, R.~D.~Ekers, E.~M.~Sadler, R.~M.~Wark, J.~L.~Higdon and E.~A.~Phinney,
  ``Radio emission from the unusual supernova 1998bw and its association with the gamma-ray burst of 25 April 1998,''
  Nature {\bf 395} (1998) 663.
 
  \bibitem{Gaisser}
Gaisser,~T.~K.,
``Cosmic Rays and Particle Physics,'' 
Cambridge and New York, Cambridge University Press, 1990, 292 p. (1990).
  
\bibitem{Wang:2007ya}
  X.~Y.~Wang, S.~Razzaque, P.~Meszaros and Z.~G.~Dai,
  ``High-energy Cosmic Rays and Neutrinos from Semi-relativistic Hypernovae,''
  Phys.\ Rev.\ D {\bf 76} (2007) 083009
  [arXiv:0705.0027 [astro-ph]].
    

\bibitem{Tacconi:2005nx}
  L.~J.~Tacconi, R.~Neri, S.~C.~Chapman, R.~Genzel, I.~Smail, R.~J.~Ivison, F.~Bertoldi and A.~Blain {\it et al.},
  ``High-resolution millimeter imaging of submillimeter galaxies,''
  Astrophys.\ J.\  {\bf 640} (2006) 228
  [astro-ph/0511319].
    
\bibitem{Hopkins:2006bw}
  A.~M.~Hopkins and J.~F.~Beacom,
  ``On the normalisation of the cosmic star formation history,''
  Astrophys.\ J.\  {\bf 651} (2006) 142
  [astro-ph/0601463].
    
\bibitem{Yuksel:2006qb}
  H.~Yuksel and M.~D.~Kistler,
  ``Enhanced cosmological GRB rates and implications for cosmogenic neutrinos,''
  Phys.\ Rev.\ D {\bf 75} (2007) 083004
  [astro-ph/0610481].

  
\bibitem{Baldry:2003xi}
  I.~K.~Baldry and K.~Glazebrook,
  ``Constraints on a universal IMF from UV to near-IR galaxy luminosity densities,''
  Astrophys.\ J.\  {\bf 593} (2003) 258
  [astro-ph/0304423].

\bibitem{Guetta:2006gq}
  D.~Guetta and M.~Della Valle,
  ``On the Rates of Gamma Ray Bursts and Type Ib/c Supernovae,''
  Astrophys.\ J.\  {\bf 657} (2007) L73
  [astro-ph/0612194].
  
\bibitem{Wanderman10}
  D.~Wanderman and T.~Piran, 
  The luminosity function and the rate of Swift's Gamma Ray Bursts
  Mon.\ Not.\ R.\ Astron.\ Soc.\  {\bf 406}, 1944 (2010) 
  [arXiv:0912.0709 [astro-ph.HE]].
  
\bibitem{Bhattacharjee:2007gc}
  P.~Bhattacharjee, S.~Chakraborty, S.~Das Gupta and K.~Kar,
  ``Upper Limit on the Cosmic Gamma-Ray Burst Rate from High Energy Diffuse Neutrino Background,''
  Phys.\ Rev.\ D {\bf 77} (2008) 043008
  [arXiv:0710.5922 [astro-ph]].
  
\bibitem{Stecker:2016fsg}
  F.~W.~Stecker, S.~T.~Scully and M.~A.~Malkan,
  ``An Empirical Determination of the Intergalactic Background Light from UV to FIR Wavelengths Using FIR Deep Galaxy Surveys and the Gamma-ray Opacity of the Universe,''
  arXiv:1605.01382 [astro-ph.HE].

\bibitem{Finke:2009xi}
  J.~D.~Finke, S.~Razzaque and C.~D.~Dermer,
  ``Modeling the Extragalactic Background Light from Stars and Dust,''
  Astrophys.\ J.\  {\bf 712} (2010) 238
  [arXiv:0905.1115 [astro-ph.HE]].
  
  
\bibitem{Berezinsky:1975zz}
  V.~S.~Berezinsky and A.~Y.~Smirnov,
  ``Cosmic neutrinos of ultra-high energies and detection possibility,''
  Astrophys.\ Space Sci.\  {\bf 32} (1975) 461.
  
  
\bibitem{Coppi:1996ze}
  P.~S.~Coppi and F.~A.~Aharonian,
  ``Constraints on the VHE emissivity of the universe from the diffuse GeV gamma-ray background,''
  Astrophys.\ J.\  {\bf 487} (1997) L9
  [astro-ph/9610176].


\bibitem{Inoue :2010}
  Inoue, Y.,
  ``High energy gamma-ray absorption and cascade emission in nearby starbust galaxies''
 Astrophys.\ J.\  {\bf 728} (2011) 11
  [arXiv:1011.6511 [astro-ph.HE]].
 
 
  
\bibitem{Lacki :2013}
  Lacki~B, Thompson~A,
  `Diffuse Hard X-ray Emission in Starburst Galaxies as Synchrotron from Very High Energy Electrons''
  Astrophys.\ J.\  {\bf 762} (2013) 29
  [arXiv:1010.3030v3 [astro-ph.HE]].
  
  
\bibitem{Chang:2014hua}
  X.~C.~Chang and X.~Y.~Wang,
  ``The diffuse gamma-ray flux associated with sub-PeV/PeV neutrinos from starburst galaxies,''
  Astrophys.\ J.\  {\bf 793} (2014) 131
  [arXiv:1406.1099 [astro-ph.HE]].



\bibitem{Torres:2004ui}
  D.~F.~Torres,
  ``Theoretical modelling of the diffuse emission of gamma-rays from extreme regions of star formation. The Case of Arp 220,''
  Astrophys.\ J.\  {\bf 617} (2004) 966
  [astro-ph/0407240].
  
\bibitem{DomingoSantamaria:2005qk}
  E.~Domingo-Santamaria and D.~F.~Torres,
  ``High energy gamma-ray emission from the starburst nucleus of NGC 253,''
  Astron.\ Astrophys.\  {\bf 444} (2005) 403
  [astro-ph/0506240].



\bibitem{Thompson:2006is}
  T.~A.~Thompson, E.~Quataert, E.~Waxman, N.~Murray and C.~L.~Martin,
  ``Magnetic fields in starburst galaxies and the origin of the fir-radio correlation,''
  Astrophys.\ J.\  {\bf 645} (2006) 186
  [astro-ph/0601626].
 
\bibitem{Lacki:2009mj}
  B.~C.~Lacki, T.~A.~Thompson and E.~Quataert,
  ``The Physics of the FIR-Radio Correlation: I. Calorimetry, Conspiracy, and Implications,''
  Astrophys.\ J.\  {\bf 717} (2010) 1
  [arXiv:0907.4161 [astro-ph.CO]].
  
\bibitem{Kneiske:2003tx}
  T.~M.~Kneiske, T.~Bretz, K.~Mannheim and D.~H.~Hartmann,
  ``Implications of cosmological gamma-ray absorption. 2. Modification of gamma-ray spectra,''
  Astron.\ Astrophys.\  {\bf 413} (2004) 807
  [astro-ph/0309141].
  
\bibitem{Gilmore:2011ks}
  R.~C.~Gilmore, R.~S.~Somerville, J.~R.~Primack and A.~Dominguez,
  ``Semi-analytic modeling of the EBL and consequences for extragalactic gamma-ray spectra,''
  Mon.\ Not.\ Roy.\ Astron.\ Soc.\  {\bf 422} (2012) 3189
  [arXiv:1104.0671 [astro-ph.CO]].
  
\bibitem{Franceschini:2008tp}
  A.~Franceschini, G.~Rodighiero and M.~Vaccari,
  ``The extragalactic optical-infrared background radiations, their time evolution and the cosmic photon-photon opacity,''
  Astron.\ Astrophys.\  {\bf 487} (2008) 837
  [arXiv:0805.1841 [astro-ph]].
  
\bibitem{Chang:2016ljk}
  X.~C.~Chang, R.~Y.~Liu and X.~Y.~Wang,
  ``How far are the sources of IceCube neutrinos? Constraints from the diffuse TeV gamma-ray background,''
  arXiv:1602.06625 [astro-ph.HE].

\bibitem{Murase:2016gly}
  K.~Murase and E.~Waxman,
  ``Constraining High-Energy Cosmic Neutrino Sources: Implications and Prospects,''
  arXiv:1607.01601 [astro-ph.HE].

  

\end{thebibliography}
\end{document}